\documentclass[prb,twocolumn,showpacs,preprintnumbers,amsmath,amssymb]{revtex4}

\usepackage{graphicx}
\usepackage{dcolumn}
\usepackage{bm}
\usepackage{subfigure}
\usepackage{float}
\usepackage{booktabs}

\newcommand{\ra}[1]{\renewcommand{\arraystretch}{#1}}

\usepackage{capt-of}

\begin{document}







\title{Fingerprints of entangled spin and orbital physics in itinerant ferromagnets \\
via angle resolved $resonant$ photoemission\\






}

\author{F. Da Pieve$^{1}$}

\affiliation{
$^1$ Laboratoire des Solides Irradi\'{e}s, UMR 7642, CNRS-CEA/DSM, \'{E}cole Polytechnique, F-91128 Palaiseau, France and European Theoretical Spectroscopy Facility (ETSF)
}

\date{\today}


\begin{abstract}




A novel method for mapping the local spin and orbital nature of the ground state of a system 
 via corresponding flip excitations in both sectors is
 proposed based on angle resolved $resonant$ photoemission and related diffraction patterns, presented here for the first time via an ab-initio {\it modified one-step theory} of photoemission. The analysis is done on the paradigmatic weak itinerant ferromagnet $bcc$ Fe, whose magnetism, seen as a correlation phenomenon given by the coexistence of localized moments and itinerant electrons, and the non-Fermi liquid behaviour at ambient and extreme conditions both remain unclear. The results offer a real space imaging of local pure spin flip and entangled spin flip-orbital flip excitations (even at energies where spin flip transitions are hidden in quasiparticle peaks) and of chiral, vortex-like wavefronts of excited electrons, 
depending on the orbital character of the bands and the direction of the local magnetic moment. Such effects, mediated by the hole polarization, 
make resonant photoemission a promising tool to perform a full tomography of the local magnetic properties of a system with a high sensitivity to localization/correlation, even in itinerant or macroscopically non magnetic systems.

\end{abstract}


\maketitle

\section{Introduction}

Spin and orbital degrees of freedom, their fluctuations, entanglement and textures, play a relevant role in many fascinating  
correlated and/or spin orbit-driven systems, like Mott insulators \cite{nagaosa,kugel,oles}, non conventional superconductors \cite{laughlin,kotliar,varma} and topological phases of quantum matter \cite{balents,zhang,natphys}. In the last two decades, it has become clear however that peculiar orbital textures and spin-orbital coupling are found even without relevant spin orbit and/or without relevant electron-electron correlation, like in low-dimensional materials exhibiting Peierls transitions and charge density waves \cite{wezel,ritschel,pabho}, in some lowly correlated insulators doped with $3d$ ions developing long range magnetic order 
\cite{dapieve}, correlated metals \cite{corrmetals} and even weak itinerant ferromagnets \cite{pou,katanin}, whose behaviour might sometimes challenge the standard model of the metallic state, the (ferromagnetic) Fermi Liquid theory. However, probing simultaneously spin and orbital degrees of freedom with high sensitivity to spatial localization is complicated, as the orbital angular momentum is often quenched by the crystal field in many relevant compounds and as the distinction between low energy spin and orbital excitations  of different nature (incoherent particle-hole and collective modes) is not always obvious \cite{minola,david}. 
Finding a strategy to improve the capabilities of widely used techniques, like angle resolved photoemission (ARPES) \cite{damascrev} and resonant inelastic X-ray scattering (RIXS) \cite{ament}, 
whose sensitivity to spatial localization is limited due to the linear dependence of the dipole operator on the spatial coordinate $\vec{r}$, would boost the advance for an atomic-scale mapping of the magnetic properties even in macroscopically non magnetic systems.

Orbital-resolved contributions to ARPES spectra are often studied either by analyzing the self-energy entering the expression of the one-body spectral function describing photoemission \cite{silke} or analyzing related dichroism signals induced by circular or linearly polarized light \cite{sanchez,boh,vee,louie}. Other more explorative works have considered Auger emission, in particular in time coincidence with photoelectrons, and unravelled the two-electron (and the corresponding two-hole) orbital contributions to both energy spectra \cite{prlstefani} and angular polar scans \cite{jes128,prbmio}. Earlier works have also studied the orbital-resolved contributions to full two-dimensional angular patterns (via the anisotropy of the excited "source wave" at the absorber) 
in core level photoemission \cite{wider,greost} and Auger spectroscopy \cite{daimon,ramak,gre}.  
The anisotropy of the charge density of such source wave(s) and the one of the core hole state (core hole polarization, $P_c$) are influenced by the polarization (and direction) of the impinging light and the polarization of the valence states. They are characterized by even multipoles (quadrupole, etc), describing the
alignment (i.e., the deviation from sphericity, given by a different occupation among the different $m_l$ states, with a symmetry between $\pm m_l$), and odd multipoles (dipole, etc), describing the
orientation  (i.e. the rotation of the charge density, given by a preferential occupation of $m_l$ states over -$m_l$ states). 

Recently, pioneering diffraction patterns have also been reported \cite{peter,magnan,greber} for resonant photoemission (RPES), the participator channel of the resonant Auger effect, the non radiative decay channel following X-ray absorption degenerate with usual ARPES. However, earlier theoretical descriptions of the resonant Auger effect, formulated on the basis of the interaction between discrete and continuum states \cite{davis}, Keldysh formalism \cite{fujikawa}, or via time-independent resonant scattering theory \cite{aberg8,gel} have not been accompanied by realistic implemented schemes. The existing, practical calculation schemes (model hamiltonian-based) \cite{tanaka,cot,degrootjes,thole} only focus on the spectator channels of the resonant Auger effect,
with two-holes-like final states, 
and not on the participator ones, where the decay occurrs before the excited electron has delocalized, leading to one-hole final states linearly dispersing with photon energy (Raman shift) , 
visible before and at the edge \cite{degrootjes}. 
Also, retriving information on local magnetic properties remains difficult, 
and some effects observed in RIXS, like spin flip-orbital flip excitations \cite{degrootjes,ament2,david,vanveenendaal,degrootprb,haverkort} have never been reported in 
RPES.

In this work, it is shown that the yet largely unexplored spin
polarized $angle$ $resolved$ RPES (AR-RPES) is a promising tool for performing a full $local$ spin and orbital tomography of the ground state of a system, by providing access to local spin flip, orbital flip and chiral excitations. 
The study is based on a recently presented ab-initio description for extended systems \cite{prlmio}, based on a $modified$ one-step theory of
photoemission, which is re-analyzed to elucidate matrix
elements effects and mixed with an auxiliary analysis of convoluted partial densities of states (DOS) to elucidate the
connection with local spin and orbital properties. 
The paradigmatic case of the weak itinerant ferromagnet bcc Fe, whose origin of ferromagnetism is nowadays seen as a correlation phenomenon, given by the coexistence of localized moments associated to electrons in a narrow $e_g$ band and itinerant electrons in the $t_{2g}$ bands, is considered. Yet unexplained correlations in the paramagnetic phase eventually determine the localization of the $e_g$ states \cite{katanin} and the formation of localized moments. Instabilities at extreme $PT$ conditions and tendency of $e_g$ states to a non-FL behaviour even for ambient conditions \cite{pou} have been reported. Analysis of ARPES spectra at different levels of theory other  than DFT (which does not contain static spin fluctuations) \cite{sanchez2} suggests the importance of non local correlations and the necessity to improve the description of (orbital-dependent) mass renormalizations. The ab-initio RPES energy spectra and diffraction patterns presented here for excitation at the $L_3$ edges by circularly polarized light show the possibility of mapping the spin polarization and local valence orbital symmetry with high sensitivity to spatial localization by analyzing spin-conserving and spin-flip exchange excitations. The results show the occurrence of pure spin flip excitations far from the Fermi level ($E_F$) and coupled spin flip-orbital flip excitations in correspondance of a narrow peak in the local partial DOS near $E_F$ associated to the elongated $e_g$ levels. 
Similarities and differences with RIXS are discussed, as well as the practical and fundamental implications concerning possible full tomographic studies of local magnetic properties and studies of spin and orbital physics in more complex systems.


\section{Theoretical Section}

The cross section for resonant photoemission is 
proportional to the Kramers-Heisenberg formula for second order processes
\begin{eqnarray}\label{eqmia1}
&&\frac{\partial^2\sigma}{\partial\Omega_p\partial\omega}\propto\nonumber\\
&&\sum_f|\langle f|D_q|0 \rangle+\sum_j\frac{\langle f|V|j\rangle\langle j|D_q|i\rangle}{E_0-E_j+i\frac{\Gamma_j}{2}}|^2 \delta(\hbar\omega+E_0-E_F)\nonumber
\end{eqnarray}

($\Gamma_j$ is the core level lifetime-induced width). 
The first term is the dipole matrix element $D_{vp}$=$\langle i\epsilon_p L_p\sigma_p|D_q|i\epsilon L_v\sigma_v\rangle$ which describes, in an effetive single particle approach, direct valence band photoemission ($v$ ($p$) denotes the valence state
(photoelectron) and $L_p=(l_p,m_p)$). The second
term represents the resonant process, described by the product of the core-absorption dipole matrix elements $D_{ck}$ and the decay (direct and exchange) matrix elements $V_d$ and $V_x$, i.e.  
$R_d=V_d \cdot D_{ck}=\langle i\epsilon_p L_p\sigma_p, j'c'|V|i\epsilon L_v\sigma_v,j'\epsilon_kL_k\sigma_k\rangle \cdot D_{ck}$
and $R_x=V_x \cdot D_{ck}=\langle j\epsilon_p L_p\sigma_p,ic'|V|j\epsilon_kL_k\sigma_p,i\epsilon L_v\sigma_v\rangle \cdot D_{ck}$ ($k$
denotes
the conduction state where the electron gets excited and $c'$ the quantum numbers $m'_c,\sigma'_c$ to
which the initial 
hole $c=m_c,\sigma_c$ might scatter). For the more localized participator decays, in the direct term the core hole is filled by the excited electron
and a valence electron is emitted, and in the exchange one the two are exchanged. In principle, the energy detuning from the
absorption edge and a narrow bandwidth of the photons can act as a
shutter between different channels, although only looking at energy spectra exhibiting the Raman shift (as often done) might not always allow the distinction between localized and delocalized excitations \cite{glatz}, which remains an open issue for both RIXS and RPES. 
All delocalized states can be described conveniently via real space multiple scattering, which 
describes the propagation of a wave in a solid as repeated scattering events \cite{seb} 
and which allows to keep explicit
dependence on the local quantum numbers.
The cross section can then be cast in a compact form as: 

\begin{equation}\label{compact}
\frac{\partial^2\sigma}{\partial\Omega_p\partial\omega}=\sum_{qq'}\varepsilon^q\varepsilon^{q'^{*}}\sigma_{qq'}\nonumber
\end{equation}

where $\varepsilon^q$ are the light polarization tensors and the hermitian 3$\times$3-matrix $\sigma_{qq'}$ is given by

\begin{eqnarray}\label{eqnmia4}
&&\sigma_{qq'} =\sum_{N,N'}K(N,q)Im\tau_v(N,N')K^*(N',q'), \nonumber\\
&&K(iL_v\sigma_v,q)=\sum_{jL_p} B^*_{jL_p}({\bf k_p})
(\delta_{ij}\delta_{\sigma_v\sigma_p}(D_{vp}+R_d)+R_x)\nonumber
\end{eqnarray}

with $N,N'$ labelling $i$ (atomic site) and $L(=l,m)$. The photoelectron scattering
amplitudes $B_{jL_p}({\bf k_p})$ can be resumed as 
$B^*_{jL_p}({\bf k_p})=Y_{L_p}({\bf k_p})i^{-l_p}e^{i\delta_{l_p}}$, i.e., 
(the source wave) + all the scattering contributions. 
The orbital and spin contribution to the outgoing electron wavefunctions are then determined by the parity and Coulomb selection rules of the whole process. They impose that 
$|l_c-|l_v-l_k||$ $\leq l_{p}\leq$ $l_c+l_v+l_k$, $l_c+l_v+l_k+l_p$=even and
$m_{c}+m_{p}=m_{v}+m_k$. 
For the spin, one has $\sigma_c=\sigma_k=\sigma_{c'}$ for the direct term (the spin of the core hole does $not$ 
flip) 
and $\sigma_c=\sigma_k=\sigma_p,\sigma_{c'}=\sigma_v$ for
the exchange term (allowing $also$ for possible core hole spin flip leading to simultaneous flip of the orbital projection $m_c$). 


The connection with ground state properties is highlighted via an auxiliary description, obtained by modifying an often used expression for normal Auger emission (i.e., a convolution of the DOS for the two final holes, \cite{fratesi}). By considering now
the DOS of the emitted electron $D(E-\epsilon)$ and the DOS of the electron dropping into the core hole $D(\epsilon)$, weighted by the core hole polarization, 
the intensity becomes:



\begin{eqnarray}\label{conv}
I_{\uparrow(\downarrow)}(E)=&& M_{\uparrow\uparrow (\downarrow\downarrow)}P_{+(-)}\int D_{\uparrow (\downarrow)}(E-\epsilon)D_{\uparrow (\downarrow)}(\epsilon)d\epsilon+\nonumber\\
&& M_{\uparrow\downarrow (\downarrow\uparrow)}P_{-(+)}\int D_{\uparrow (\downarrow)}(E-\epsilon)D_{\downarrow (\uparrow)}
(\epsilon)d\epsilon \nonumber
\end{eqnarray}


where $P_{\pm}=(1{\pm}P_{c})/2$ takes into account the modifications of the DOS of the electron filling the hole by the core hole polarization, and $M_{\uparrow\uparrow (\downarrow\downarrow)}$ and $M_{\uparrow\downarrow (\downarrow\uparrow)}$ are respectively the sum of the modulus squares of the spin conserving (direct and exchange) decay matrix elements $and$ the modulus square of the spin flip (exchange) decay matrix element:

\begin{eqnarray}\label{convM}
&&M_{\uparrow\uparrow (\downarrow\downarrow)}= |V_{d,\uparrow\uparrow (\downarrow\downarrow)}|^2+|V_{x,\uparrow\uparrow (\downarrow\downarrow)}|^2,\nonumber\\
&&M_{\uparrow\downarrow (\downarrow\uparrow)}= |V_{x,\uparrow\downarrow (\downarrow\uparrow)}|^2\nonumber
\end{eqnarray}

$P_c$ is in a range from -1 \cite{sinko} (as in a ferromagnet with spin down holes, and light impinging parallel to the magnetization), to some other values $<1$ when the hole flips or the photon polarization and the local valence polarization form a generic angle (in this latter case, both even and odd
multipoles contribute to $P_{c}$ \cite{kab}, and dicroism occurrs in both 
absorption and decay). 


The important theoretical prediction can
then be made that the occurrence of spin flip transitions and their
entanglement with orbital ones are determined by the (geometry-dependent) core hole
polarization, how it affects excited states of different degree of localization/delocalization, and how it weights the decay
exchange matrix elements. Also, orbital flips should be more visible when perturbing a highly symmetric 
(with respect to relevant quantization axis) intermediate-state orbital population (alignment), rather than an asymmetric one. Given the influence of matrix elements on
different allowed source waves and the high energy of the
photoelectrons (which reduces the importance of final-state
effects), it can be expected that a selective real-space mapping
of (local) spin and spin-orbital excitations is
possible by looking at two-dimensional angular patterns. 




\section{Computational details}

\begin{figure*}[!htb]
\begin{center}
\includegraphics[clip=,width=16.0cm,height=8.0cm]{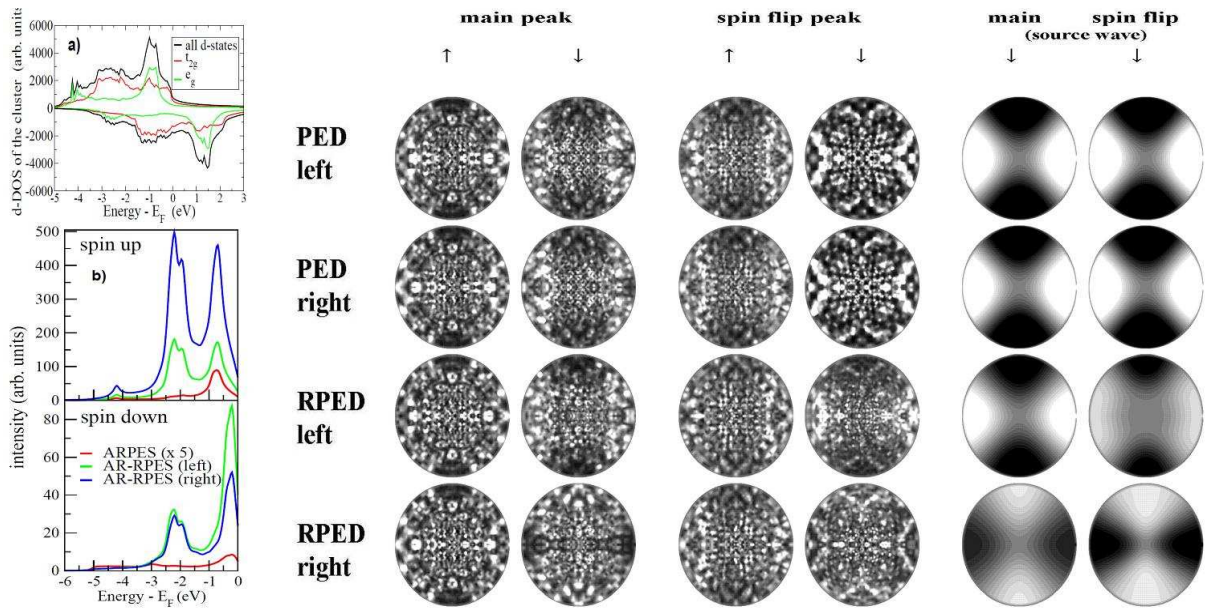}
\end{center}
\caption{a) DOS of the Fe(010) cluster; b) ARPES and AR-RPES spectra 
   (from \cite{prbmiorpes}) for parallel geometry and normal emission. Rest of the panel: PED, RPED  
for initial binding energy corresponding to the main peak and the spin flip peak in the spin up AR-RPES spectrum, 
and ``source waves'' patterns (the emitter is embedded in the cluster but no scattering events take place). 
(the plotted 
function is $\chi=I[\theta,\phi,\epsilon]/I_0[\theta,\epsilon] -1$, 
$I_0$ being the intensity averaged over all $\phi$-dependent values. Scans are around the surface normal.)}
\label{fig:geom1}
\end{figure*}

Excitation at the 
$2p_{\frac{3}{2}}$ edge of the itinerant weak ferromagnet Fe by circularly polarized light is investigated
to proof the unique capabilities of RPES. 
A semispherical Fe(010) cluster (with 184 atoms and in-plane magnetization along $<$001$>$), and  
DFT spin polarized potentials obtained by a scalar relativistic LMTO  \cite{andersen} calculation for bulk Fe bcc in DFT-local spin density approximation (LSDA)  
are used as input for a multiple scattering code developed by the author, which can calculate usual ARPES and RPES
 from cluster type objects. The spectra and full hemispherical patterns are obtained taking into account the interference due to emission from different atomic sites when exciting at resonance. 
The calculated magnetic moment of 2.26 $\mu_{\rm B}$ from the self-consistent calculation is in good agreement 
with experiment. Core states are calculated atomically by solving the Dirac equation, while delocalized states (bound and unbound) are developed, as mentioned before, via multiple scattering. The photoelectron is described as a time-reversed LEED state, i.e. a plane wave with linear momentum {\bf k} plus incoming spherical waves on all atoms. A real inner potential (10 eV) is used which serves as a reference energy
inside the solid with respect to the vacuum level and inelastic damping is included via a constant imaginary potential (4.5 eV). For the optical transitions, the dipole approximation in the              
acceleration form is used, since the length form is 
not well defined for delocalized state. The weak spin-orbit (SO) coupling of the valence and continuum states
has been neglected.

From a theoretical viewpoint, non radiative decays are complicated dynamical processes which include atomic relaxation and electron screening in response to the core hole. However, reasonable approximations can be made for Fe.
Electron-core hole interaction is generally weak in metals because of efficient screening of the Coulomb interaction and its only observable effect is the $reduced$ branching ratio between the $L_2$ and $L_2$ edges of the isotropic x-ray absorption spectra, with respect to what obtained within the independent particle approximation. However, such reduction is generally smaller for spin-polarized and dichroic spectra, and more importantly, in RPES it only affects the intermediate states, which are not directly observed. For Fe, the deviation of the branching ratio from the statistical value is actually very small \cite{ebert}, indicating a reasonable description in terms of a single particle approach. 
Also, as a consequence of being a weak ferromagnet, both minority and majority spin states can be populated to screen the core hole, leading to no drastic change in the local moment \cite{renn}. When the decay takes
place, with a valence electron filling the hole and the excited electron emitted, either the effective potential seen by the valence electrons is restored to its initial form or, as the electron is emitted with high kinetic energy, a sudden  response of the valence electrons occurrs due to the destruction of the core hole, with no time for electrons to readjust. Thus the spin polarization of the emitted electron results to be approximately the one of the intermediate state, very similar though, for Fe, to the one of the initial ground state \cite{nohltig}. Dipole and Auger-like matrix elements are then calculated here using ground state scalar relativistic wave functions.
The robustness of the approach is demonstrated by earlier successful comparisons between calculated spin polarization, energy spectra and photoemission diffraction patterns and experiments \cite{prlmio,prbcu}.

\section{Results}

Fig.~\ref{fig:geom1}a, 1b show the $d$-DOS of the $whole$ cluster and the ARPES and AR-RPES spectra for a photon
energy at the maximum of the resonance for normal emission and
$parallel$ geometry (light impinging along the magnetization, along which spin is measured). 
The ARPES spectra show each one
main peak, 
absence of other sharp features as for a genuine lowly correlated system, in agreement with experiments
\cite{surfsci}, and null dichroism, due to non chiral geometry and neglected SO in delocalized states. In contrast, the resonant spectra
exhibit dichroism (in this geometry only related to the  
absorption step as the orientation of the core hole is unaffected by reversal of helicity \cite{corehole}) and, more importantly, new peaks. 
Going towards higher binding energies, the spin up RPES spectra show a first (second) peak for emission from $e_{g}^{\uparrow}$ ($t_{2g}^{\uparrow}$) states, while the spin down
spectra exhibit a first peak for emission from $t_{2g}^{\downarrow}$ states and then an unexpected peak 
at an energy where
there are almost no spin down states in the DOS, and which thus corresponds to spin up valence
states. This means that the spin of the photoelectron is
opposite to the one of the final valence hole, 
and thus it is a spin
flip transition. Such (exchange-induced) spin flip can only occur for $2p_{3/2}$ eigenstates with $mixed$ spin
character due to SO 
(the $m_j=\pm 1/2$ sublevels, 
$|3/2,1/2(-1/2)>=\sqrt{2/3}|Y_{10}^{\uparrow}(Y_{10}^{\downarrow})>+\sqrt{1/3}|Y_{11}^{\downarrow}(Y_{1-1}^{\uparrow})>$). 

We now move to the more explorative resonant diffraction patterns.
Ab-initio spin polarized resonant and direct photoemission diffraction
patterns (RPED, PED) are reported 
in Fig.~\ref{fig:geom1}, for initial energies corresponding to the two peaks in the $spin$ $up$ AR-RPES spectra (the main peak near $E_F$ and the one at higher binding energy, corresponding to the spin flip excitations in the spin down channel). 
It is clear that, while almost all RPED patterns resemble the corresponding direct ones,  
a net 
90$^o$ twist occurrs for right circular polarization for the RPED pattern of the spin down channel, the one 
allowing for spin flip transitions, a clear signature
of an accompanying $orbital$ flip of the photoelectron wave. 
Interestingly, the effect is actually mainly visible at the
main peak, revealing 
spin flip transitions hidden by dominating spin-conserving ones in the quasiparticle peak.

\begin{table}\centering
\ra{1.3}
\caption{\label{tab:table1} Exchange transitions at core states with mixed spin character, for left (right) polarization $\Delta m=+1(-1)$. 
}
\begin{ruledtabular}
\begin{tabular}{llcrrrr}
$\Delta m$ & $edge$ & $m_c;\sigma_c$ & $m_k;\sigma_k$ & $m'_c;\sigma'_c$ & $m_p;\sigma_p$ & $m_v;\sigma_v$ \\
\colrule
+1 & $\frac{3}{2}$;-$\frac{1}{2}$ & 0;-$\frac{1}{2}$ & 1;-$\frac{1}{2}$  & -1;$\frac{1}{2}$ & {\bf 3},4,{\bf 2},{\bf 1},0;$-\frac{1}{2}$ & {\bf 1},2,{\bf 0},{\bf -1},-2;$\frac{1}{2}$  \\
+1 & $\frac{3}{2}$;$\frac{1}{2}$ & 1;-$\frac{1}{2}$ & 2;-$\frac{1}{2}$ & 0;$\frac{1}{2}$ & 3,4,2,1,0;$-\frac{1}{2}$ & 1,2,0,-1,-2;$\frac{1}{2}$ \\
-1 & $\frac{3}{2}$;-$\frac{1}{2}$ & 0;-$\frac{1}{2}$ & -1;-$\frac{1}{2}$ & -1;$\frac{1}{2}$ & {\bf 1},2,{\bf 0},{\bf -1},-2;$-\frac{1}{2}$ & {\bf 1},2,{\bf 0},{\bf -1},-2;$\frac{1}{2}$ \\
-1 & $\frac{3}{2}$;$\frac{1}{2}$ & 1;-$\frac{1}{2}$ & 0;-$\frac{1}{2}$ & 0;$\frac{1}{2}$ & {\bf 1},2,{\bf 0},{\bf -1},-2;$-\frac{1}{2}$ & {\bf 1},2,{\bf 0},{\bf -1},-2;$\frac{1}{2}$\\

\end{tabular}
\end{ruledtabular}
\end{table}

This orbital flip phenomenon can be understood via the two models described in the theoretical section, by analyzing the exchange matrix elements and the local partial DOS.
The selection rules dictate $l_{p}$=1,3,5 (with 3 numerically found as the most
probable wave, 
in line with previous works on similar transitions \cite{gre,term}).
Table I reports the exchange transitions occurring at core
hole states with $mixed$ spin
character (at their spin down
components,  
as available empty states are spin down core hole states will also be mainly spin down). 
These are $mixed$ spin flip-orbital flip
transitions, in which 
both the $m_l$ and $\sigma_z$ components of the $same$
$m_j$ substate flip. 
Transitions mixing different $m_j$s, like $m_j=1/2$ flipping to $m_j=-1/2$, are 
also possible, being the $m_j$ sublevels separated by 0.32 eV, 
but these imply only spin flip.
We recall that the relevant  
irreducible representations here are: $t_{2g}$: $d_{xy}=\frac{1}{\sqrt{2}}(\psi_2-\psi_{-2}),d_{yz}=\frac{1}{\sqrt{2}}(\psi_1-\psi_{-1}),d_{zx}=\frac{1}{\sqrt{2}}(\psi_1+\psi_{-1})$; $e_g$: $d_{x^2-y^2}=\frac{1}{\sqrt{2}}(\psi_2+\psi_{-2}),d_{3z^2-r^2}=\psi_0$. Their contribution to the partial DOS around a central absorber ion in the the cluster is shown in Fig. 2.


\begin{figure}
\begin{center}
\includegraphics[clip=,height=4.0cm,width=0.35\textwidth]{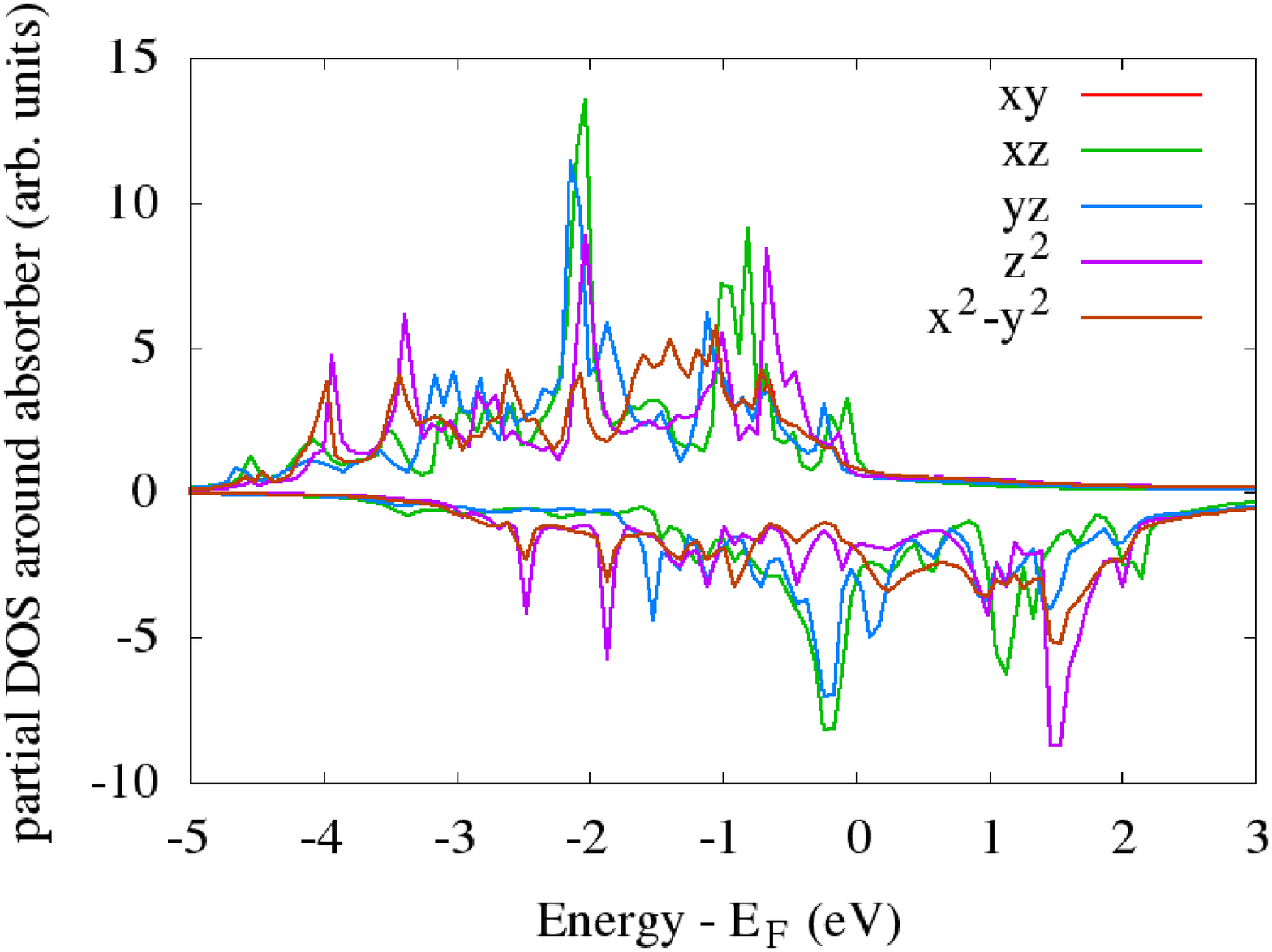}
\end{center}
\caption{Local partial DOS ($l,m$- resolved) around a Fe central ion in the cluster}
\label{fig:dospartabs}
\end{figure}

For left-handed light ($\Delta m=+1$ here), the excitation to a $m_k=1,\downarrow$ state ($t_{2g}^{\downarrow}$) (first row in Table I) is more probable than
photoexcitation of the other spin down component of the other sublevel \cite{corehole}. The numerical evaluation of the decay matrix elements   
for different orbital contributions, similarly to earlier investigations \cite{apecsCu,epl2011,renn}, allows to select the
dominant transitions (in bold in Table I),
and it partially reflects the reasonable result that the
decay is more favourable if the two involved valence and conduction electrons 
have the maximum number of equal quantum numbers, as in this case they will repeal more. 
The decay leading to a $t_{2g}^\uparrow$ final hole with
$m_v=\pm 1$ ($d_{xz}$,$d_{yz}$) gives the strongest contribution, making a distinction between different orbitals in the DOS around the absorber ion. Indeed, considering the localized nature of the
recombination, such DOS unravels the
orbital character of the decaying states 
better than 
the DOS of the
whole cluster, 
revealing narrow and prounounced 
peaks from different orbitals of the two irreducible
representations in the spin up main peak, reminding of Van Hove singularities 
in the extended electronic structure \cite{maglic,irkhin}.
Angular momentum conservation rules then dictate a 
$Y_{33}^\downarrow$ emitted wave, with strong intensity reduction along the
quantization axis, similarly to the one expected in direct photoemission from a $d$-shell (Clebsch-Gordan coefficients
indeed give the highest probablity for a final $m=\pm$3 state
generated by photon absorption at the $m_v=\pm$2 states) 
and in line with previous reports on aligned $f_{\pm3}$
emitted waves for different compounds \cite{ramak}. 
For right-handed light ($\Delta m=-1$), the absorption is equally probable at
the two spin down components of the two mixed spin
character \cite{corehole} sublevels. However, again the numerical evaluation of 
the product of the matrix elements suggests distinct contributions to the decay, notably a decreasing contribution from the $d_{xz}$ valence states and a stronger one from the $e_{g}^{\uparrow}$ states with
$m_v=0$ ($d_{3z^2-1}$). 
This leads to a $\sim Y_{30}^\downarrow$ emitted wave, indeed twisted by $90^o$ with respect to the
$\sim Y_{3\pm3}$ behaviour expected in usual photoemission by left/right polarization. 
At the spin flip energy, the effect seems absent, due to a stronger $e_{g}^{\uparrow}$-$t_{2g}^{\uparrow}$ hybridization and the contribution from more than one orbital of the
same irreducible representation (the
$d_{xz},d_{yz}$ orbitals of the $t_{2g}^{\uparrow}$).
This leads to more balanced
contributions of $m_l$ waves and to a petal-like structure. 

The results are the first demonstration that RPES is sensitive to the very orbital nature of the ground state, as for elongated orbitals ($d_{3z^2-1}$) a different type of spin-flip transition (mixed with an orbital flip) is allowed, contrary to the planar $x^2-y^2$ and interaxial $t_{2g}$ orbitals, similarly to what previously observed in RIXS \cite{marra}. The phenomenon indeed reminds of the (local) orbital excitations (local $dd$ excitations) 
often studied by RIXS via changes in the polarization of the scattered light. Here such excitations manifest themselves as deviations from the anisotropy expected in usual photoemission and can accompany spin flip satellites in the spectra, even when hidden in the quasiparticle peak. 
Contrary to ARPES,  
the photoelectron wave then reflects exactly the orbital character of the 
valence state, 
allowing to map the valence orbital
symmetries via monitoring the angular
distribution of the resonant current of opposite spin.


For the aim of accessing correlated orbitals and understanding the very nature of their resonant excitations, an important observation has to be done: the (exchange-induced) spin flip-orbital flip excitations involve an $e_g^{\uparrow}$ hole which, being in a completely filled majority spin band, is
more localized than those in the partially filled
minority spin. These more localized valence flip excitations are then transferred to the photoelectron. The visible orbital flip effect is thus a manifestation of a different correlation in the two bands with  different spin, established recently on a quantitative basis by experimental and theoretical studies on Auger emission \cite{prlstefani}, and of different orbital character, as earlier suggested \cite{goodenough}.
Orbitals appear nearly as quenched far from $E_F$, where only spin flip excitations are clear, while spin and orbital degrees of freedom are entangled and both active at low energy.

This has three fundamental implications. First, 
it is relevant to underline that, at least in the normal Auger decay, spin flip transitions are not expected to remember of
the photon angular momentum in a two step process and should be
$always$ balanced by an orbital flip to conserve the total
angular momentum $\Delta J_z=0$ due to the scalar nature of the Coulomb interaction. The results here suggest that, at resonance and in a one-step approach, spin flip transitions might not be always accompanied by orbital flip (as it occurrs at the energy of the spin flip satellite) and that, even when occurring with orbital flip, as in correspondance of the elongated (and more localized) $d_{3z^2-1}$, there is a memory on the photon polarization. This suggests that both the Raman shift and the possible memory on the polarization as seen in the angular distributions should be considered when trying to make a distinction between localized and delocalized excitations. Second, despite the local crystal field description used here, the results suggest that in a general more complex superexchange scenario, the counterpart collective excitations (magnons and orbital waves) might also be accessed. This however would require a mapping of two-dimensional patterns for different detuning energies from the resonance, such to distinguish incoherent particle-hole excitations from collective modes via their dependence/independence on the photon energy \cite{minola}. Third, the observed entangled spin-orbital physics in the $e_g$ band of Fe due to enhanced correlations suggests that precursor traces of the non-Fermi liquid behaviour observed at extreme $PT$ \cite{katanin} and ambient \cite{pou} conditions can be traced even in the phase of ideal $PT$ conditions, often though of insignificant correlations. Notably, the entangled spin and orbital degrees of freedom get active at the narrow $e_g$ peak near $E_F$, reminiscent of a Van Hove singularity \cite{maglic,irkhin} in the electronic structure, indeed earlier invoked to be partially responsable for the above mentioned instabilities.

At last, an important practical implication is brought by the fact that the flip effect has an atomic nature, as shown by the spin down source waves patterns (Fig.~\ref{fig:geom1}),     
and it disappears for the spin unpolarized phase (Fig.3). This demonstrates the sensitivity of RPES to spatial localization, due to the dominance of on-site
transitions \cite{prlmio} caused by the 1/$r$ behaviour of the Coulomb operator and by the localization of the excited core orbital, 
opening the path for elementally sensitive imaging of magnetic domains. Practical implementations might well involve cutting-edge techniques such as spectromicroscopy \cite{asensio}, with energy, angle and high lateral resolution, opening the route for magnetic tomographic photoemission. 

\begin{figure}
\begin{center}
\hspace*{2.0em}\includegraphics[clip=,width=3.6cm,height=0.7cm]{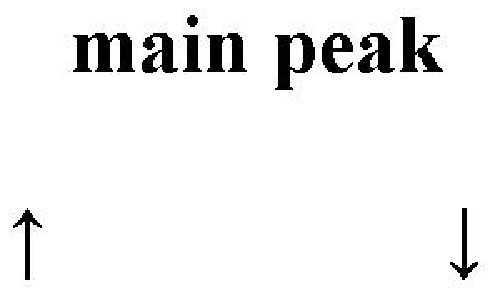}\\
\end{center}
\parbox{0.08\columnwidth}{
\begin{center}
\hspace*{-1.0em}\includegraphics[clip=,width=1.2cm,height=6.6cm]{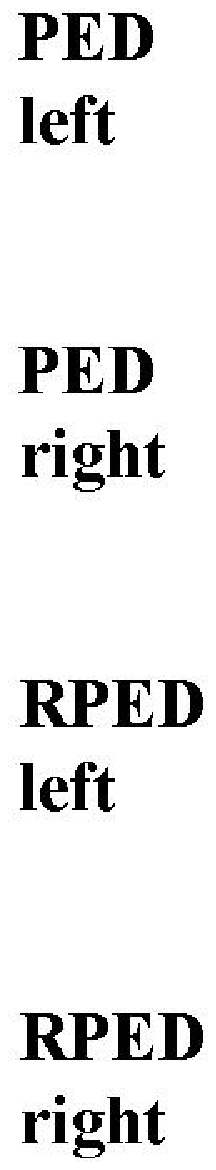}
\end{center}
}
\parbox{0.4\columnwidth}{
\begin{center}
\hspace*{1.3em}\includegraphics[clip=,height=1.7cm,width=0.09\textwidth]{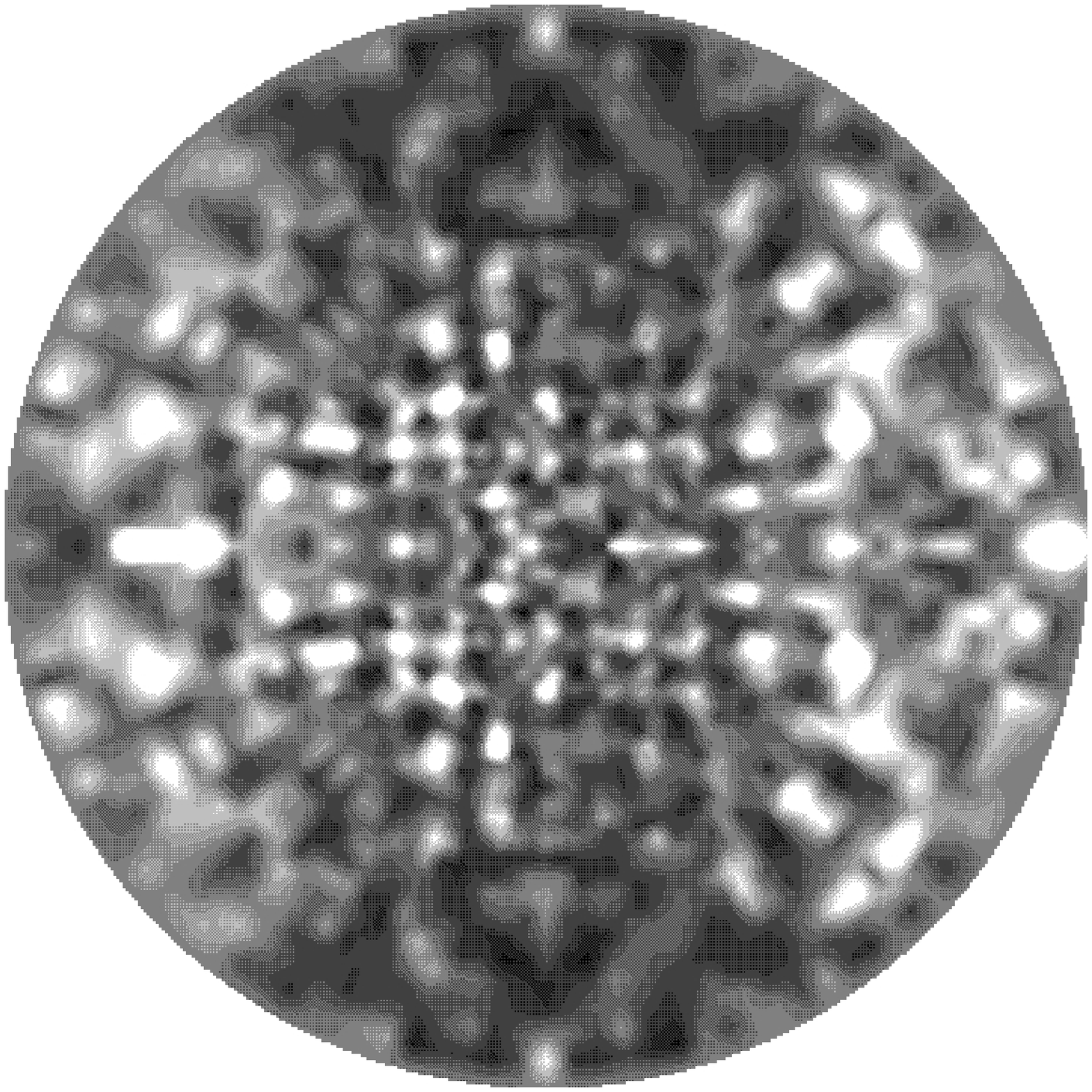}\\
\hspace*{01.3em}\includegraphics[clip=,height=1.7cm,width=0.09\textwidth]{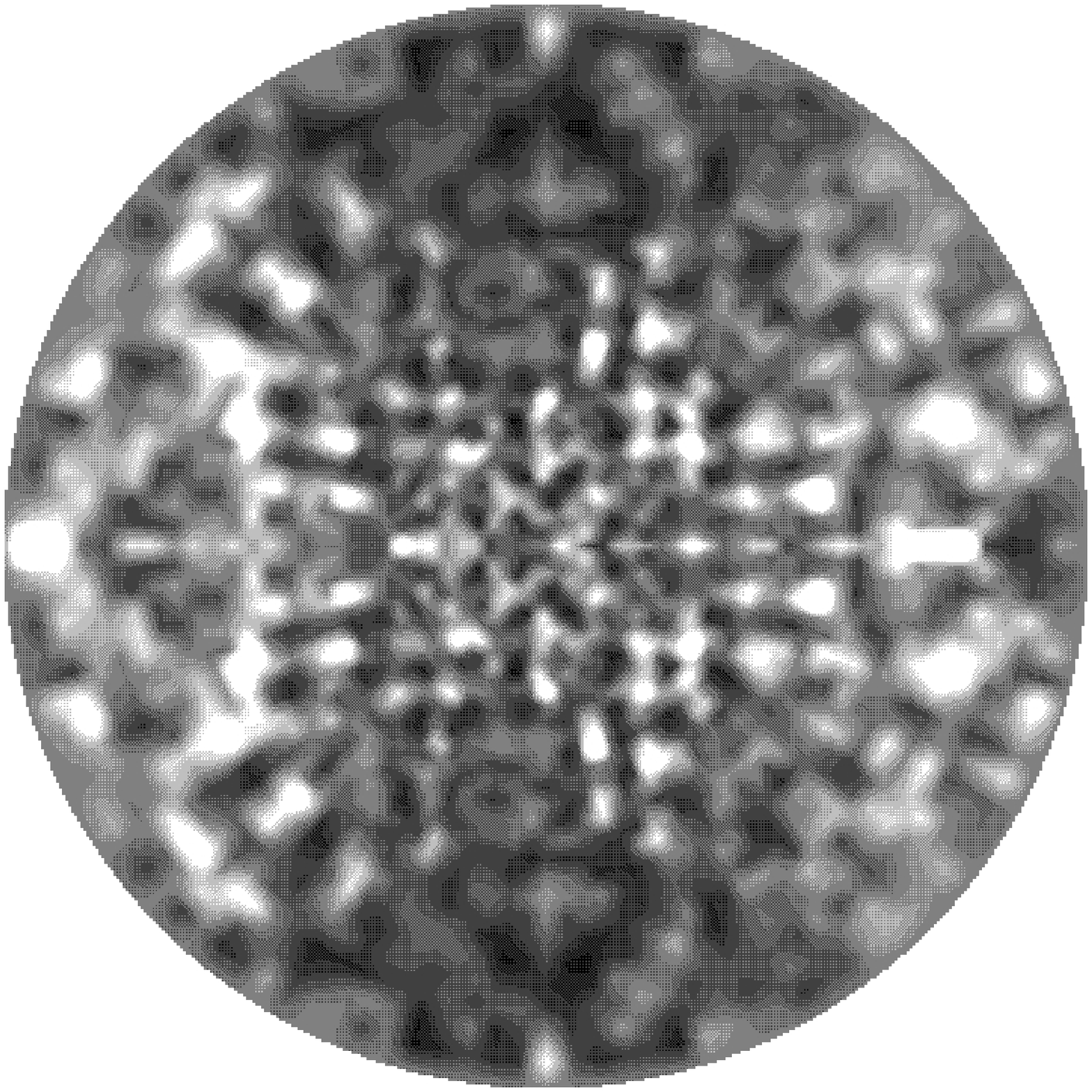}\\
\hspace*{1.3em}\includegraphics[clip=,height=1.7cm,width=0.09\textwidth]{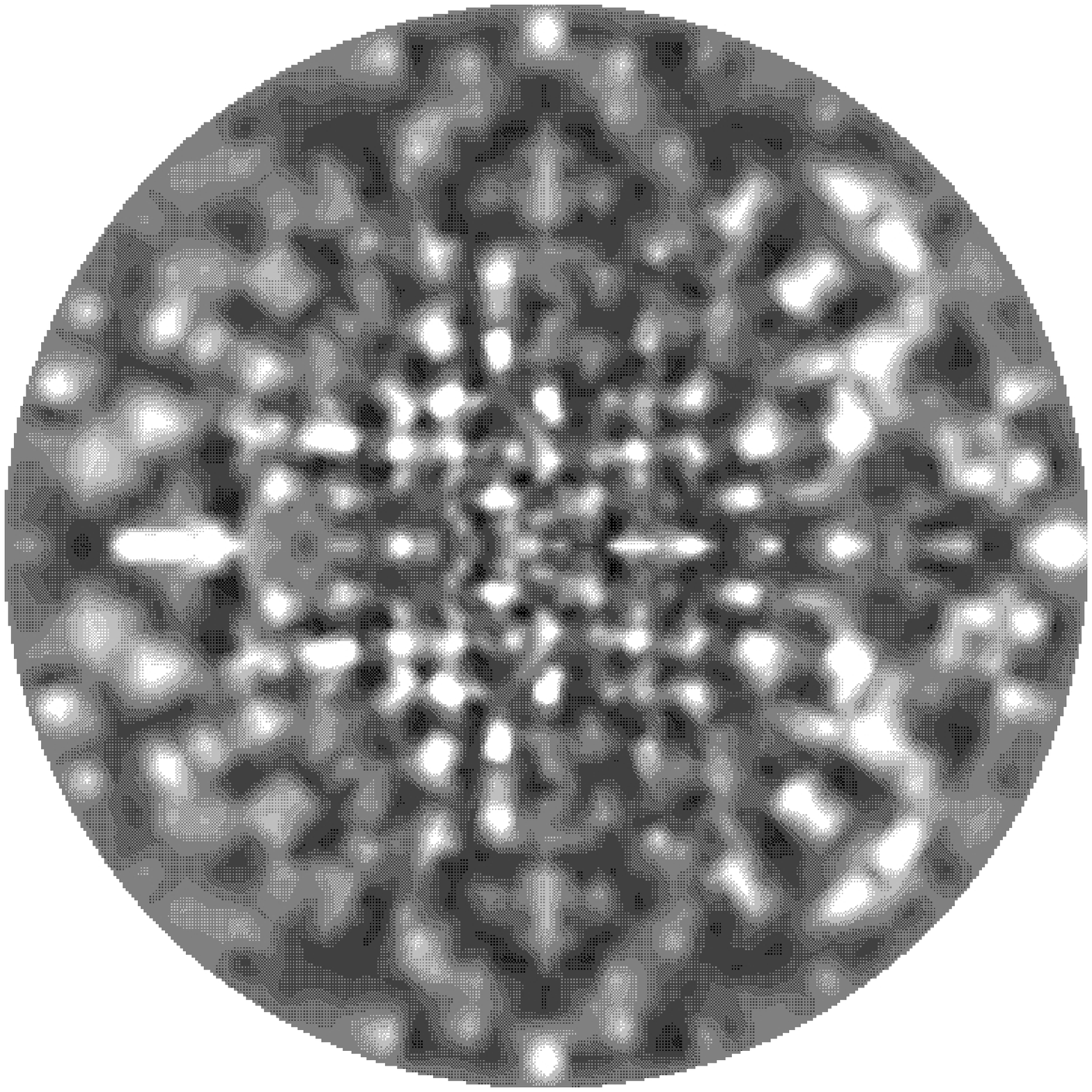}\\
\hspace*{1.3em}\includegraphics[clip=,height=1.7cm,width=0.09\textwidth]{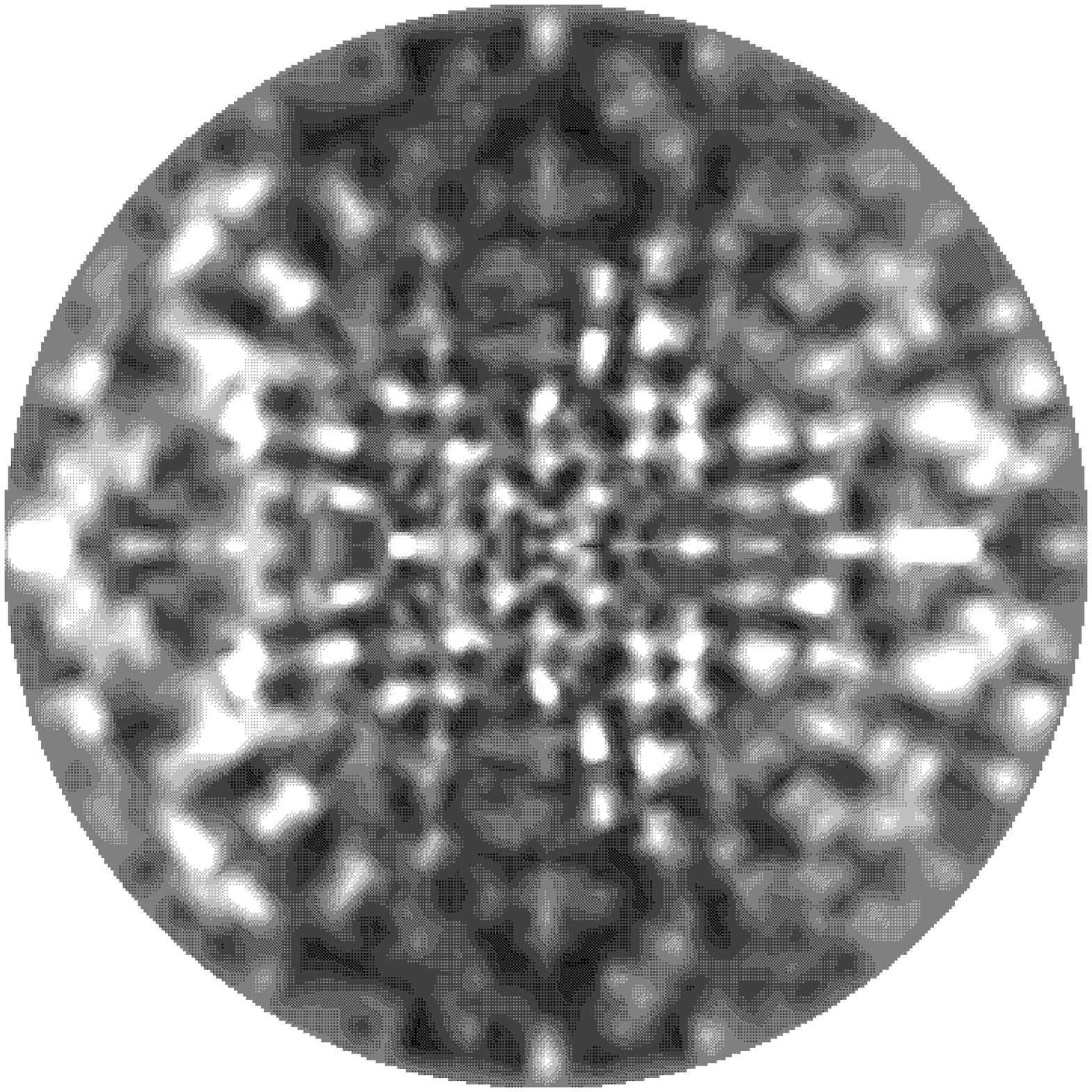}
\end{center}
}
\parbox{0.4\columnwidth}{
\begin{center}
\hspace*{-1.5em}\includegraphics[clip=,height=1.7cm,width=0.09\textwidth]{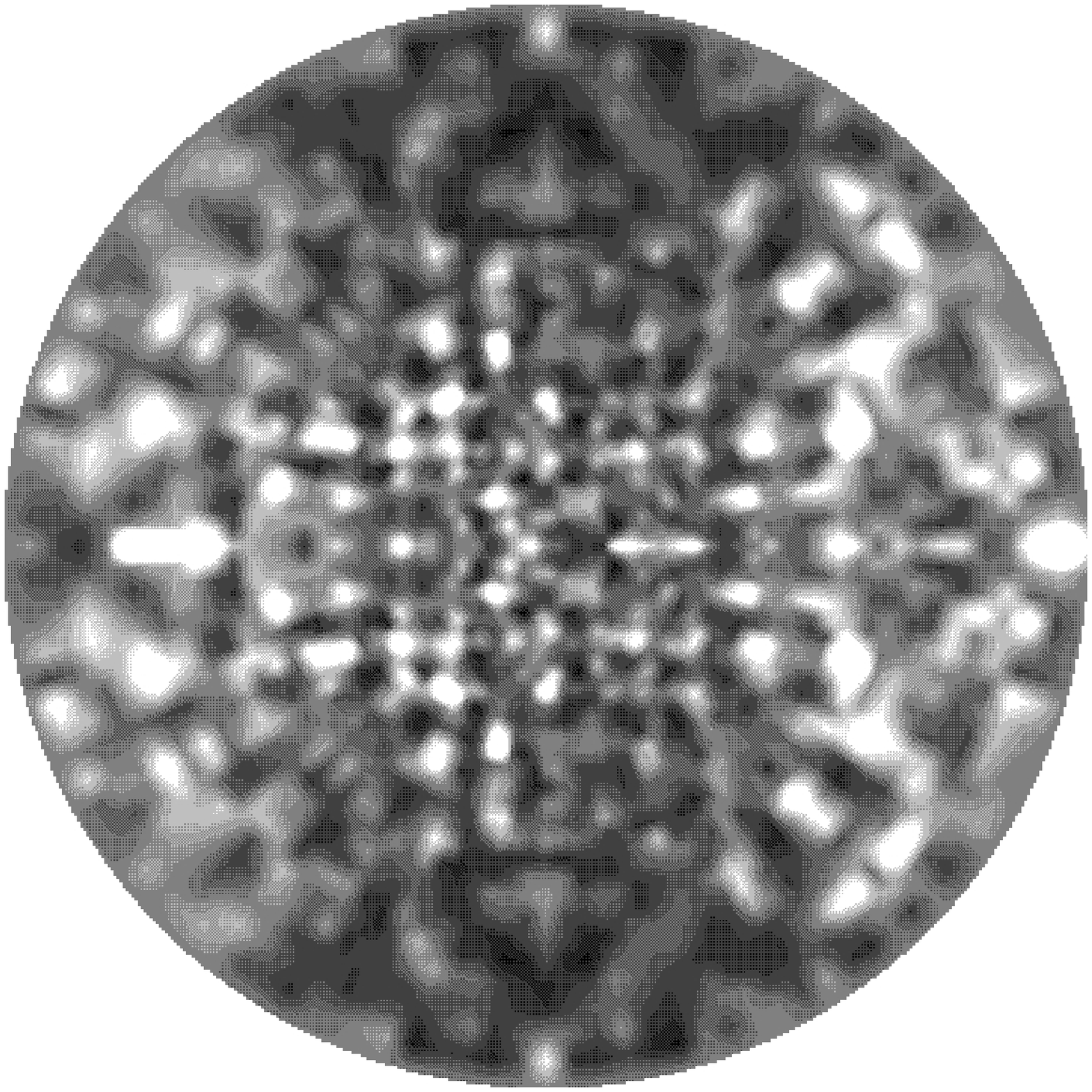}\\
\hspace*{-1.5em}\includegraphics[clip=,height=1.7cm,width=0.09\textwidth]{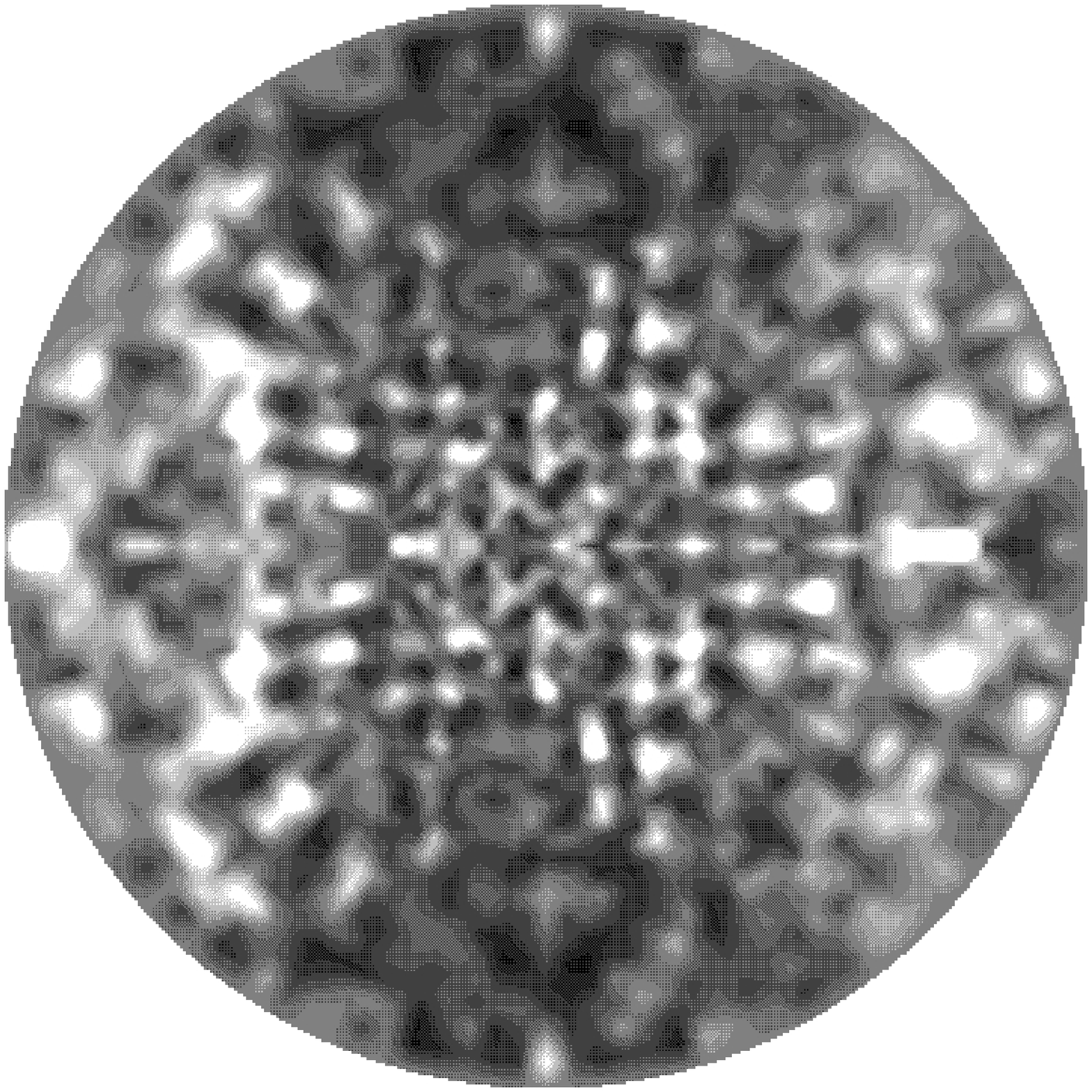}\\
\hspace*{-1.5em}\includegraphics[clip=,height=1.7cm,width=0.09\textwidth]{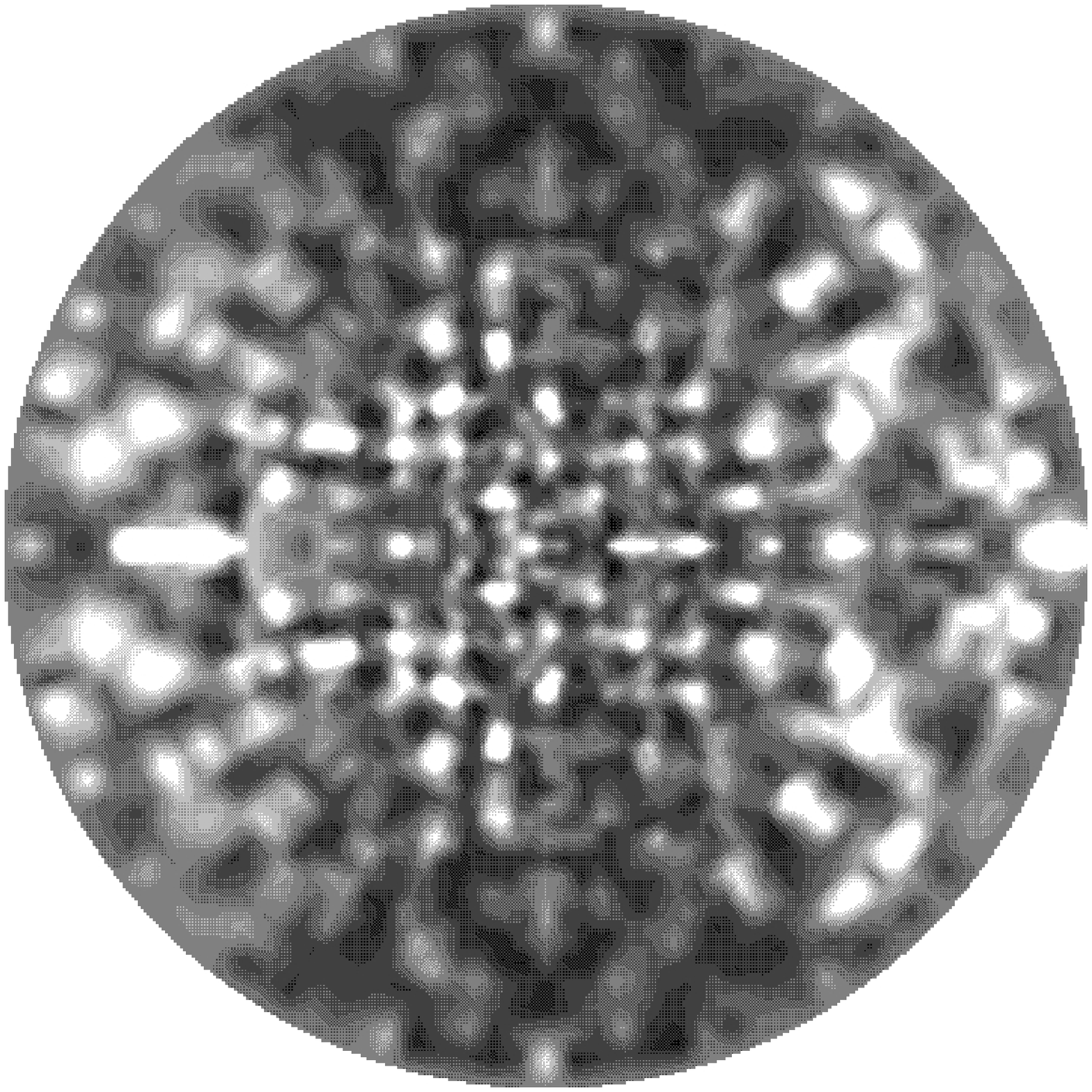}\\
\hspace*{-1.5em}\includegraphics[clip=,height=1.7cm,width=0.09\textwidth]{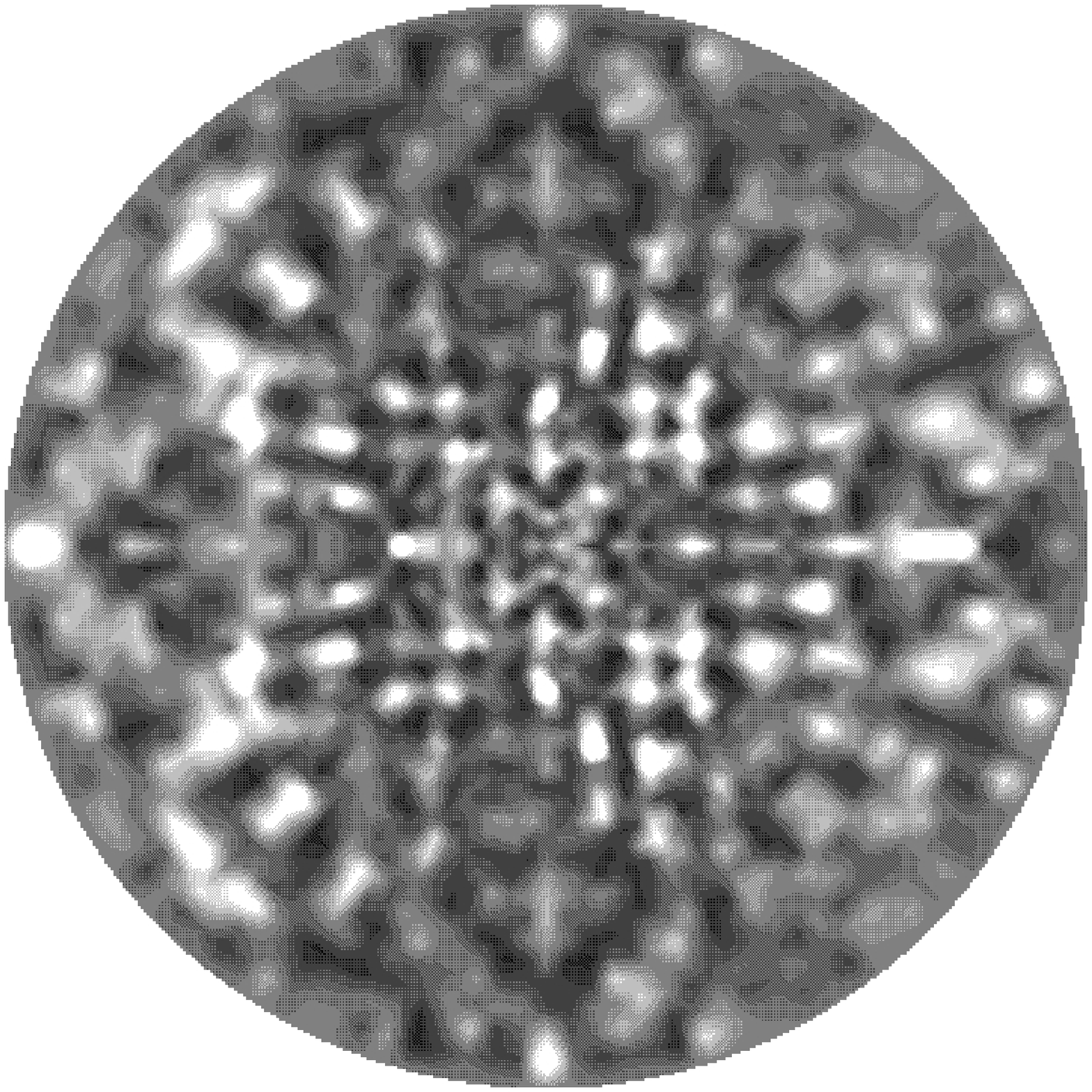}
\end{center}
}
\caption{Spin polarized PED and RPED patterns for parallel geometry,
  for excitation at the $L_3$ edge for paramagnetic Fe(010), photon energy at the maximum of the resonance and initial state energy corresponding to the main peak in the spin up channel for the ferromagnetic phase.}
\label{fig:gePMparalL3}
\end{figure}

\begin{figure}[!htb]
\begin{center}
\includegraphics[clip=,height=7.0cm,width=8.9cm]{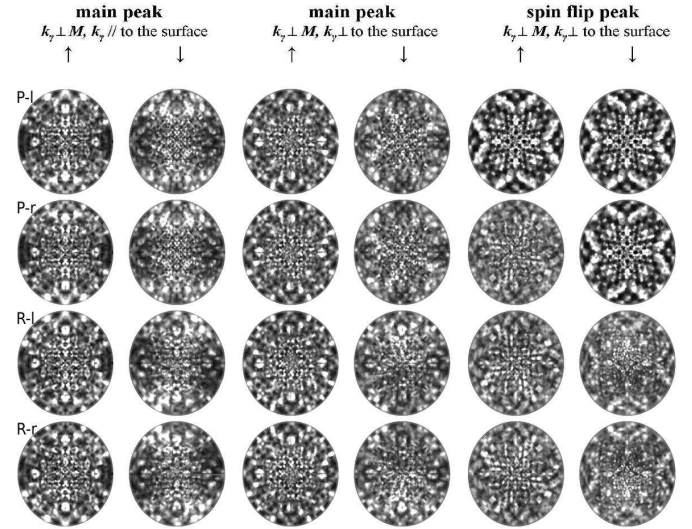}
\end{center}
\caption{PED (P) and RPED (R) patterns for two perpendicular geometries for left (-l) and right (-r) polarization.}
 \label{fig:second}
\end{figure}

The situation changes drastically when the core hole polarization changes, i.e. when the photon helicity and the local magnetic moment are oriented differently.
Fig.~\ref{fig:second} reports the patterns for $two$
$different$ perpendicular geometries (light impinging
perpendicularly to the magnetization), 
for which the dichroism in absorption is nulla but the core hole polarization (now both the deviation from sphericity of the charge density and its rotation) 
does influence differently the emission for left and right handed light.
As the incident light direction is rotated away from the
quantization axis, the selection rules will actually now allow a mixture
of $\Delta m=0,\pm 1$ transitions and thus a detailed microscopic analysis of orbital contributions is more complicated. However, some clear fetures can be observed. 
For grazing incidence, 
(only the main peak energy is considered), 
the spin down RPED patterns again deviate from the direct ones, 
and exhibit a rotation between the two polarizations, though different from the previous 90$^o$ flip. 
Interestingly, when the light is impinging
$perpendicularly$ to the surface, 
and thus the scan around the surface normal coincides with a scan around the photon incidence direction, 
vortex-like features, given by crosses of higher intensity with bending arms following the counterclockwise (clockwise) rotation of the electric field
for left (right) handed light, appear for specific channels.
Such effect, called
circular dichroism in angular distributions 
and previously
observed in direct photoemission even from non magnetic and non chiral structures 
\cite{daimon,daimon2,fadley}, is due to forward scattering peak "rotations" related to the $m_l$ of the emitted wave, and is 
here unveiled to be correlated with local valence orbital symmetries. Emission from the $t_{2g}$ (spin down (up) emission for the main (spin flip) peak energy), differentiating from the $e_g$ states by non isotropic
combinations of $m_l$s, can easily favour non balanced
combinations with preference towards $\pm m_l$ in the continuum wave,
according to photon's helicity. 
Chirality in the patterns thus remains, as the emitted wave
is now oriented (the asymmetries do not cancel when summing
over its $m_l$ components).
At the spin flip energy, the spin down channel corresponds to emission from mixed $e_g$-$t_{2g}$ states, and again a petal-like pattern appears. 
For the resonant patterns, orbital twists are weakened or absent, suggesting smaller contributions of spin flip terms and a delocalized valence hole. 



\section{Conclusions}

In summary, this work  
presents the exciting prospect of a new generation of resonant photoemission experiments, 
capable to probe 
simultaneously the spin polarization, the (energy resolved) local valence orbital symmetries and the orientation of local magnetic moments,
exploiting 
the core hole polarization as
a prism to access spin and orbital excitations. 

The results suggest that the combined analysis of angle-resolved resonant photoemission energy spectra and diffraction patterns 
can give profund insights into the physics of many fascinating materials. In case of Fe, a coupling between spin and orbital degrees of freedom near the Fermi level is reported, suggesting it as crucial element in the developement of a unified theory of magnetism encompassing both the localized moments and the itinerant behaviour picture for this system. 
The access to the corresponding different excitations according to the local orbital symmetry and degree of localization would allow for example to probe metal-oxygen and metal-metal orbital hybridizations for different energies in oxides, and to probe the competition between electron localization and delocalization in Mott insulators and correlated metals. The work obviously also suggests that matrix elements effects have to be considered in the description of resonant photoemission, which necessarily has to go beyond interpretations based on sole spectral functions or estimations of matrix elements averaged over the full valence region. Last, the results also challenge the more conventional use RIXS to probe spin and orbital physics, opening the doors for a possible exploration of both incoherent particle-hole and collective magnetic excitations also via the non radiative resonant channel.

\begin{acknowledgments} 
The author acknowledges fruitful discussions with P. Kr\"{u}ger at the very early stage of this work, the financial support from the EU (Marie Curie
Fellowship, FP7/2007-2013, Proposal No 627569) and the COST action MP1306: Modern Tools for Spectroscopy on Advanced Materials.
\end{acknowledgments}

\bibliography{achemso-demo}

\end{document}